\documentclass[iop,twocolappendix,floatfix]{emulateapj}

\usepackage{lineno}
\usepackage{epsfig}
\usepackage{apjfonts}
\usepackage{amsmath,color,bm}
\usepackage{booktabs}
\usepackage[breaklinks,colorlinks,urlcolor=blue,citecolor=blue,linkcolor=blue]{hyperref}
\usepackage[utf8]{inputenc}
\usepackage[T1]{fontenc}

\shorttitle{Constraining compact dark matter from strong lensing of GWs}
\shortauthors{Barsode et al}

\usepackage{natbib}
\usepackage{nameref}
\usepackage{graphicx}
\usepackage{graphics}
\usepackage[space]{grffile}
\usepackage{latexsym}
\usepackage{amsfonts,amsmath,amssymb}
\usepackage{url}
\usepackage[utf8]{inputenc}
\usepackage{fancyref}
\usepackage{hyperref}
\usepackage{xcolor}
\usepackage[normalem]{ulem}

\def\zl{{z_\ell}}

\def\dTl{{\Delta T_\ell}}

\def\Ml{{M_\ell}}

\def\fdm{f_\mathrm{DM}}

\def\Msun{{\mathrm{M}_{\odot}}}
\def\fdm{{f_\mathrm{DM}}}

\def\NL{{N_\ell}}

\begin{document}

%\title{Constraining compact dark matter from the non-observation of strong lensing of gravitational waves}
\title{Constraints on compact dark matter from the non-observation of gravitational-wave strong lensing}
\author{A. Barsode$^1$}
\author{S. J. Kapadia$^{2,1}$}
\author{P. Ajith$^{1,3}$}
\affiliation{$^1$~International Centre for Theoretical Sciences, Tata Institute of Fundamental Research, Bangalore 560089, India}
\affiliation{$^2$~Inter-University Centre for Astronomy and Astrophysics, Pune 411007, India}
\affiliation{$^3$~Canadian Institute for Advanced Research, CIFAR Azrieli Global Scholar, MaRS Centre, West Tower, 661 University Ave., Suite 505, Toronto, ON M5G 1M1, Canada}

\begin{abstract}
    We use the non-observation of strong lensing of gravitational waves (GWs) in the first three observation runs of LIGO-Virgo detectors to constrain the fraction of dark matter in the form of compact objects in the mass range $10^{6}-10^{9}~{\Msun}$. Using a Bayesian formalism supplemented by astrophysical simulations of strong lensing of GWs, we constrain the compact dark matter fraction to $\lesssim 0.4 - 0.6$ with currently available data and show that they may get significantly tighter in the future. We find that multiple lensing --- i.e. GWs getting deflected by multiple compact objects on their way to us --- is possible. By ignoring this, we underestimate the constraints by a few percent.
\end{abstract}

\keywords{strong lensing, gravitational waves, dark matter, compact objects, multiple lensing}
%\maketitle

\section{Introduction}
\label{sec:intro}
Gravitational wave (GW) astronomy has emerged as a powerful cosmological probe that complements electromagnetic (EM) observations. The weak interaction of GWs with matter renders them exceptionally well-suited for delving into the dark universe. So far, the ground-based interferometric detectors LIGO and Virgo have detected over 90 GW events in their first three observing runs \citep{abbott2019gwtc, abbott2021gwtc, abbott2023gwtc, abbott2024gwtc}. Most of these have been identified as binary black hole (BBH) coalescences, with a few as binary neutron star (BNS) and neutron star-black hole (NSBH) coalescences.

Because the coupling between matter and gravity is weak, GWs undergo minimal attenuation on their way through the cosmos. However, like light, their paths may get distorted due to the presence of intervening matter, through the effect of gravitational lensing. Galaxies and galaxy clusters are expected to be the most promising lenses. Since the characteristic length scale ($\sim$ Schwarzschild radius, $G \Ml/c^2 \sim 10^{12}$ km) of these objects is much larger than the wavelength of GWs ($\lambda_\mathrm{GW} \sim 10^{3}$ km), their lensing can be treated in the geometric optics approximation. Theoretical studies of lensing of GWs due to galaxies have suggested that $\sim 0.1-1\%$ of all GW signals observable by the first three observation runs of LIGO-Virgo detectors may be strongly lensed, producing multiple images \citep{ng2018precise, li2018gravitational, oguri2018effect}. The expected lensing probability due to clusters is likely an order of magnitude smaller~\citep{smith2017strong}.

Compact objects (COs) like black holes (BHs) could also produce lensing effects on GWs. If the gravitational radius of the lensing CO is comparable to the GW wavelength ($G \Ml/c^2 \sim \lambda_\mathrm{GW}$), then geometric optics approximation becomes inaccurate and lensing will involve wave optics effects (diffraction, interference, etc). In the frequency band of ground-based GW observatories ($f_\mathrm{GW} \sim 10-10^3~\mathrm{Hz}$), CO lenses in the mass range $\sim 10^2-10^5~M_\odot$ are expected to produce wave optics effects (called, GW microlensing)~\citep{takahashi2003wave}, while more massive CO lenses ($\Ml \gtrsim 10^5~M_\odot$) can produce multiple images (strong lensing), similar to galaxies. The probability of observing such effects depends on the abundance of such COs in the universe. The expected abundance of astrophysical BHs (produced from the collapse of baryonic matter) is not large enough to produce observable lensing signatures in the near future. However, if a significant fraction of dark matter is in the form of primordial BHs or other exotic COs, their lensing effects will be observable soon.

Several searches performed in LIGO-Virgo data failed to obtain significant evidence of wave optics effects or multiple images in the $\sim 90$ GW events observed so far \citep{hannuksela2019search, mcisaac2020search, abbott2023search, LIGOScientific:2021izm, janquart2023follow, goyal2024rapid}. While this non-observation is so far not in conflict with the expectations for standard lens populations, it can be used to rule out some more exotic proposals. For example, the non-observation of microlensing-induced distortions in the GW signals has been used to constrain the fraction of dark matter in the form of COs in the mass range $10^2-10^5~\Msun$~\citep{basak2022constraints}. While their constraints are modest, it is nevertheless a novel and independent demonstration of using GWs to probe the nature of dark matter.

In this paper, we constrain the fraction $\fdm \equiv \Omega_\mathrm{CO}/\Omega_\mathrm{DM}$ of dark matter in the form of COs in the mass range $10^{6}-10^{9}~\Msun$ from the non-observation of strong lensing of GWs. Here, $\Omega_\mathrm{DM}$ is the energy density of the dark matter and $\Omega_\mathrm{CO}$ is that in the form of COs. While the condition that GWs be strongly lensed imposes the lower limit of $\sim 10^5~\Msun$, there is no upper bound on the masses of objects that may be constrained. Yet, it may not make sense to go beyond the masses of the smallest dwarf galaxies ($\sim 10^9~\Msun$).

Constraints have already been put on this range of CO masses \citep{carr2021constraints}; the tightest ones arising from the non-observation of distortions in CMB spectrum ($\fdm < 10^{-6}$ for $\Ml > 10~\Msun$). However, these are model dependent and may be evaded if the COs are presumed to be born lighter and allowed to accreate matter \citep{carr2022primordial}. The most robust constraints come from the non-observation of dynamical effects these COs would have on various structures in the universe ($\fdm < 10^{-4}$ for $\Ml \sim 10^7~\Msun$) \citep{carr1999dynamical}. Non-observation of strong lensing of EM sources has also been used for constraining $\fdm \lesssim 10^{-2}$ \citep{PhysRevLett.86.584, zhou2022constraints, nemiroff2001limits, ji2018strong, leung2022constraining, PhysRevLett.121.141101}.

In comparison, we constrain $\fdm$ to be less than $0.4-0.6$ at 90\% credible interval using currently available data. While these constraints are quite modest, this is the first demonstration of a new probe of compact dark matter in this mass range using GWs. Further, we shall show that these constraints will get significantly better in the next few years as the number of GW detections grows rapidly.

\section{Strong lensing by compact dark matter}
\label{sec:strong-lensing}

\subsection{The strong lensing optical depth}
\label{sec:strong_lensing_optical_depth}

In order to constrain the fraction $\fdm$ of dark matter in the form of COs, we need to compute the relationship of $\fdm$ with the detectable fraction of GW events lensed by such objects. The primary quantity bridging these two is the strong lensing optical depth $\tau(z_s)$ --- the average number of lenses whose area of influence (described by the Einstein radius) intersects the line of sight to a source at redshift $z_s$:
\begin{equation}
\tau(z_s) = \int_0^{z_s} \dfrac{d\tau}{d\zl}\left(\zl, z_s \right),
\label{eq:opt_depth}
\end{equation}
where ${d\tau}/{d\zl}$  is the differential optical depth contributed by lenses at redshift $\zl$. If we assume that these COs are uniformly distributed in comoving volume, the differential optical depth is given by~(see, e.g., \cite{jung2019gravitational})
\begin{equation}
\dfrac{d\tau}{d\zl}\left(\zl, z_s \right) = \fdm\Omega_\mathrm{DM} \dfrac{3H_0^2}{2c} \dfrac{\left(1 + \zl\right)^2}{H(\zl)} \dfrac{D(\zl) D(\zl, z_s)}{D(z_s)}.
\label{eq:diff_opt_depth}
\end{equation}
Here, $D(\zl), D(z_s)$ and $D(\zl, z_s)$ denote the angular diameter distance to the lens, to the source and between the lens and the source, $H(z)$ is the Hubble parameter with $H_0$ being its current value, $\Omega_\mathrm{DM}$ is the energy density of dark matter in units of the critical energy density of the universe, and $c$ is the speed of light.

If one assumes that dark matter is made up of COs, the propagation of waves no longer follows the null geodesics of the smooth universe. A simple way to understand this is by considering the waves propagating through a light cone to us from a source. If the source is not lensed, then, by definition, there is no CO inside this light cone, and therefore, no dark matter. Waves would travel differently in such an ``empty cone'' than if it had been filled with a smooth distribution of dark matter. Detailed calculations presented in \cite{nikolaev2015development} and \cite{nikolaev2016effect} show that the angular diameter distance in an empty cone cosmology can be larger than the standard ``filled cone'' cosmology. As a result, the optical depth is smaller in an empty cone (lower panel of Fig.~\ref{fig:z_distrib_optical_depth}).

The empty cone cosmology refers only to the case $\fdm=1$. For $0<\fdm<1$, the cone may be ``partially'' filled. In this case, the optical depth is between that of the empty and completely filled case. However, it is difficult to incorporate this effect in our simulations, so we resort to using the empty cone assumption for all $\fdm$. This would underestimate the optical depth, and therefore the lensing fraction, and consequently make our $\fdm$ constraints more conservative. For the background cosmology, we use the \textsc{Planck18} \citep{aghanim2020planck} parameter values: $h=0.674,\ \Omega_{m}=0.315$.

\subsection{The case for multiple lensing}

The optical depth definition in Eqs.~(\ref{eq:opt_depth}--\ref{eq:diff_opt_depth}) considers sources falling inside the Einstein radius $r_E$ of the lenses ($y \leq 1$, where $y$ is the impact parameter in units of $r_E$).  However, for a single point lens, two images will be formed even when the sources are outside the Einstein radius (see, e.g., \cite{meneghetti2021introduction}), though magnification of the second image rapidly decreases when $y\gtrsim 1$ (Fig.~\ref{fig:point_lens_mag}). The average number of lenses having an impact parameter less than $y$ is given by $y^2 \, \tau(z_s)$.

Usually, a fixed upper cutoff on $y$ is chosen, with the expectation that all the sufficiently magnified lensed events will be taken into consideration because any lensed event with an impact parameter greater than $y_0$, though more likely to occur, will not be magnified enough to be detectable.

% --------------------------------------------------------------------------------------------------------------------------------------------- %
\begin{figure}[t]
    \centering
    \includegraphics[width=\columnwidth]{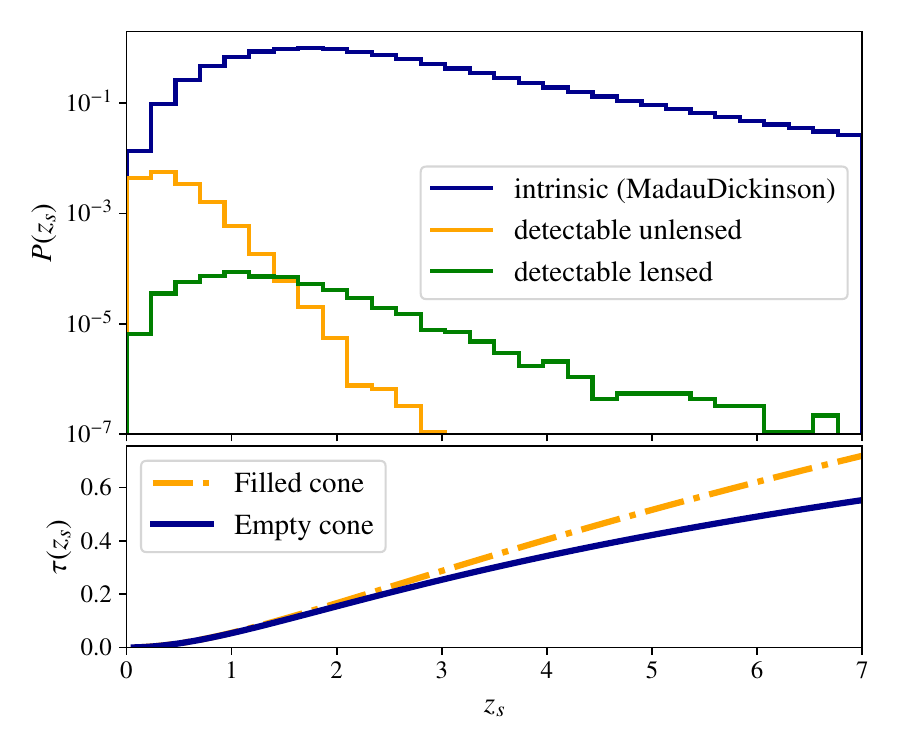}
    \caption{\textit{Upper panel}: The intrinsic distribution of the redshifts of BBHs, as well as the distribution of detected BBHs (both unlensed and lensed -- these are scaled to reflect their abundance relative to the intrinsic population). Please refer to Appendix~\ref{sec:simulations} for details of our simulations. \textit{Lower panel}: Strong lensing optical depth for the smooth (filled cone) and inhomogenous (empty cone) cosmologies.}
    \label{fig:z_distrib_optical_depth}
\end{figure}
% --------------------------------------------------------------------------------------------------------------------------------------------- %

The important caveat is that the above expectation is valid only when the optical depth is small. If it isn't, there is a good chance that GWs may encounter multiple lenses along the way ($y_0^2\tau(z_s) > 1$). Almost all events will be lensed (the lensing probability is $P_\ell(z_s)=1-e^{-y_0^2 \tau(z_s)}$), and their magnifications may no longer follow the same behavior as for the single lens case. As the upper panel of Fig.~\ref{fig:z_distrib_optical_depth} shows, the horizon for detecting strongly lensed events in LIGO-Virgo detectors is deep enough that we must include all the dark matter up to $z=7$ as potential lenses, leading to very high optical depths. Thus, it would appear that we must include the effects of multiple lensing in our analysis of strong lensing by compact dark matter.

Dealing with multiple lensing is complicated due to the intrinsic non-linearity of the lens equation (see, e.g.,~\cite{schneider1992gravitational}). The simplest way to deal with multiple lensing is to consider only the effect of the strongest lens among the lenses encountered by a source~\citep{peacock1986flux}. We show in Appendix \ref{sec:multiple_lensing_approx} that this will result in a more conservative estimation (by $< 2.5\%$) of the fraction of lensed events from our simulations.  Keeping in mind, therefore, that our constraints will be more conservative, we proceed with using the strongest lens approximation.

\section{Has LIGO-Virgo detected multiple images due to compact dark matter lenses?}
\label{sec:non-observation}

Using the strongest lens approximation, we can simulate lensing of GWs and obtain a relationship between $\fdm$ and the detectable lensing fraction $u(\fdm, \Ml)$, for COs of mass $\Ml$. If the expected lensing fraction is higher and if we observe no lensed events, constraints on $\fdm$ will be tighter. In this section, we argue that though we cannot be absolutely certain about the non-detection of strong lensing in LIGO-Virgo data, we can be reasonably sure that we have not detected strong lensing due to COs in the mass range $10^{6}-10^{9} \Msun$.

\begin{figure}[t]
    \centering
    \includegraphics[width=\columnwidth]{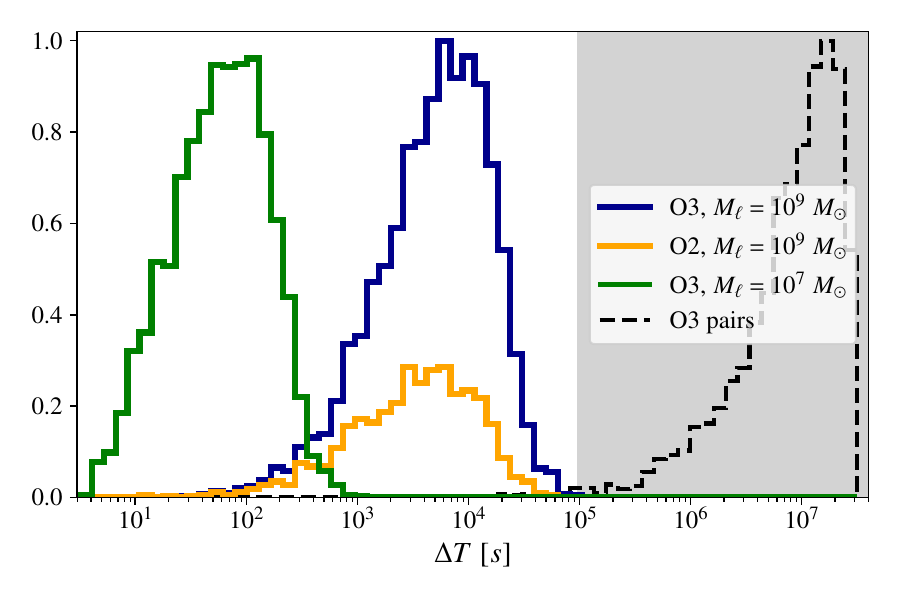}
    \caption{The solid lines show the distribution of time delays between detectable strongly lensed image pairs as obtained from our simulation (scale relative to the ``O3, $10^9 \Msun$'' case). The assumed observing run and the mass of the lensing CO are shown in the legend.  The dashed line shows the distribution of time delays between pairs of events observed in O3 (scaled to peak at 1). Even if all of the dark matter was made of $10^{9}~\Msun$ COs, the lensing time delay would not exceed $\sim 10^5$ s (gray shaded region). We have used a GW source redshift distribution from \cite{madau2014cosmic} for this simulation.}
    \label{fig:timedelay_distribs}
\end{figure}

Figure~\ref{fig:timedelay_distribs} shows the \emph{expected} distribution of the time delays between lensed images obtained from simulations (see Appendix \ref{sec:simulations} for details). We find that a multiply imaged event pair that is strongly lensed by COs of mass $10^{6}-10^{9}~\Msun$ cannot have a time delay $\gtrsim 10^5$ s. On the other hand, all the events in the first two observing runs (O1 and O2) of LIGO-Virgo have time delays of at least a few days \citep{abbott2019gwtc, abbott2021gwtc}, and therefore, none of them could be pairs of images lensed by COs. In the third observing run (O3), all except 23 pairs have a time delay $\gtrsim 10^5$ s \citep{abbott2023gwtc, abbott2024gwtc}. Therefore, these pairs could not have been lensed by COs. Similarly, we rule out the top 5 candidates of the subthreshold lensing search by \cite{abbott2023search} since all of these have time delays greater than a day.

The 23 pairs which \textit{do} have time delay smaller than $\sim 10^5$ s have little consistency in their sky locations and other inferred source parameters. Their strong lensing likelihood ratios as calculated using the \emph{posterior overlap analysis} \citep{haris2018identifying} are $\leq 1.8\times 10^{-2}$ \citep{ligo_scientific_collaboration_and_virgo_2024_10841987}. This likelihood ratio (Bayes factor) quantifies the overlap integral between the posterior distributions of the parameters corresponding to these events. It is highly improbable for truly lensed pairs to have Bayes factors as small as $\sim 10^{-2}$~\citep{haris2018identifying}. The 24th pair in increasing order of time delays, \textsc{GW200219\_094415 - GW200220\_124850}, has a Bayes factor of $21.9$~\citep{ligo_scientific_collaboration_and_virgo_2024_10841987}, which is perhaps too high to be dismissed as unlensed. In fact, we \textit{chose} $10^9 \Msun$ as the upper limit of CO mass range so that the time delays caused by them would not exceed this pair's time delay (97475 s) and we can be certain that we have not detected any lensing due to COs of this mass.

Before proceeding, we briefly mention how we could still put constraints on $\fdm$ if we observe a pair of events having a time delay smaller than what a CO lens might give. In such a scenario, the probability mass contained below this observed time delay in the simulated time delay distribution is still not observed. However, the probability mass contained above the observed time delay has to be down-weighted because of our inability to rule out lensing. The weight factor is simply given by the probability of getting a lensed pair of events from CO lenses having a Bayes factor smaller than the Bayes factor of the observed pair.

\section{Constraining $\fdm$ from non-observation of strong lensing}
\label{sec:constraints}

To use the non-detection of strongly lensed GWs to constrain the fraction of compact dark matter, $\fdm$, we use the same Bayesian formalism as that used in \cite{basak2022constraints}. We assume that the total number of detected GW events and the number of lensed events follow Poisson distributions with means $\Lambda$ and $\Lambda_\ell$, respectively\footnote{The point mass lenses will create two images. We call an event as lensed only when both of these are detectable. Note, also, that the lensed events may come from a larger distance as compared to their unlensed counterparts, owing to the lensing magnification (see the top panel of Fig.~\ref{fig:z_distrib_optical_depth}). We account for this in our astrophysical simulations.}. The detectable lensing fraction is defined as a ratio of the Poisson means of the numbers of lensed and total events detected, $u\equiv \Lambda_\ell / \Lambda$. A posterior on $u$ can be obtained using the fact that we have observed 90 events so far in O1, O2, and O3, and none of them were lensed (please refer to \S3 of \cite{basak2022constraints} for a full derivation).
\begin{equation}
\label{eq:u_posterior}
P(u \mid N, \NL=0) \propto \int\limits_0^\infty d\Lambda \, \Lambda^{N+1} e^{-\Lambda(u+1)} \, \mathcal{P}_\ell(u\Lambda) \, \mathcal{P}(\Lambda)
\end{equation}
Here, $N(=90)$ is the total number of detected GW signals, $\NL(=0)$ is the number of strongly lensed events among them, $\mathcal{P}_\ell$ and $\mathcal{P}$ are the assumed priors for the lensed and total event rates respectively. The proportionality constant is fixed by normalizing $P(u \mid N, \NL=0)$.

\begin{figure*}[t]
    \centering
    \includegraphics[width=2\columnwidth]{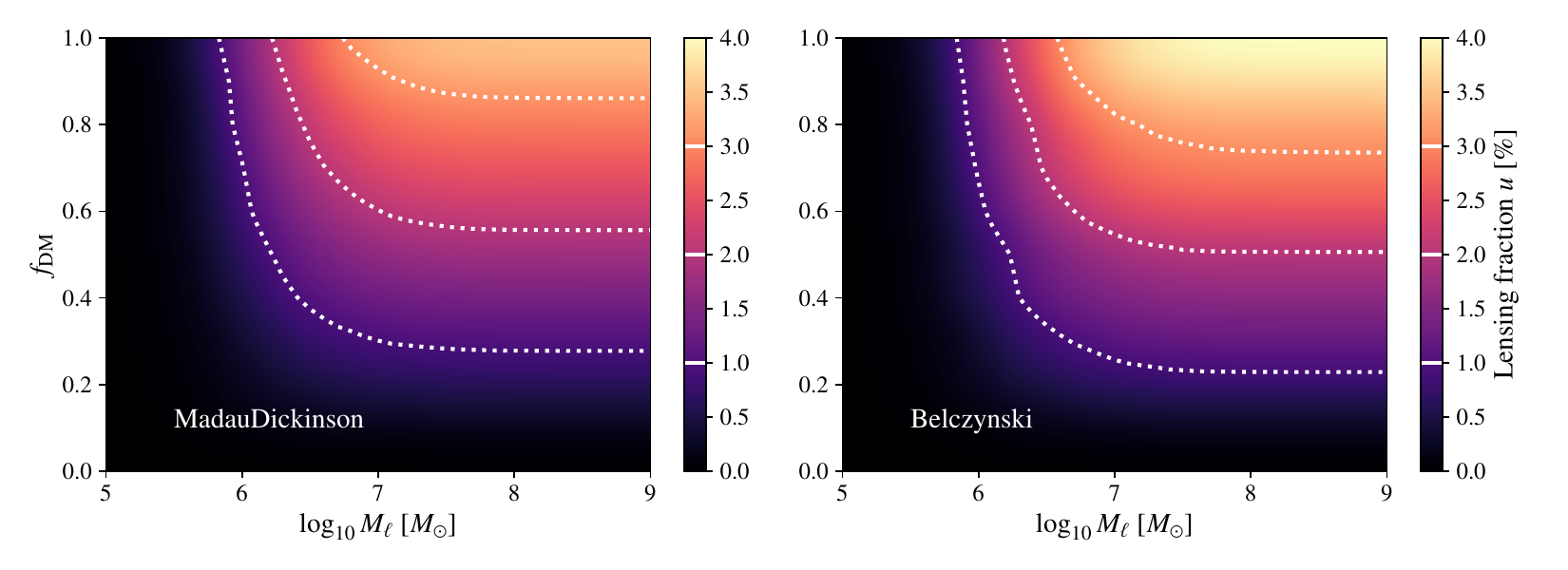}
    \caption{Expected fraction $u$ of detectable strongly lensed events if a fraction $\fdm$ of dark matter is made up of COs of mass $\Ml$.}
    \label{fig:u_combined}
\end{figure*}

Next, we simulate strong lensing by compact dark matter to estimate the relationship between the expected lensing fraction $u$ versus the fraction of dark matter $\fdm$, for a given mass $\Ml$ of the COs. The details of our simulation are given in Appendix~\ref{sec:simulations}. The final result of these simulations, a map of $u$ as a function of $\fdm$ and $\Ml$, is shown in Fig. \ref{fig:u_combined} for two choices of GW source redshift distribution.

% ------------------------------------------------------------------------------------------------------------ %
% ------------------------------------------------------------------------------------------------------------ %

Overall, the lensing fractions are higher for \cite{belczynski2016first}'s redshift distribution, which has a higher merger rate at higher redshifts than \cite{madau2014cosmic}'s. This is a direct consequence of the fact that the optical depth is higher at higher redshifts, leading to higher chances of lensing. Since the probability of encountering lenses increases with $\fdm$, it is natural to expect $u$ to increase monotonically with $\fdm$.

More interesting is the behavior of $u$ with $\Ml$: for small $\Ml$, there are very few detectable events with time delays larger than the signal duration so that they may appear as distinct strongly lensed images. Consequently, the lensing fraction is nearly zero for $\Ml \lesssim 10^6~\Msun$. For a point lens, the lensing time delay is directly proportional to $\Ml$, leading to a steady growth in the number of events having time delay greater than the signal duration. Eventually, the fraction saturates near $\Ml \sim 10^7~\Msun$ as all the detectable events have sufficiently large time delays.

With the relationship between $u$ and $\fdm$ thus known, we take its numerical derivative to estimate the Jacobian required to transform the posterior on $u$ (Eq.~\ref{eq:u_posterior}) to that on $\fdm$ \citep{basak2022constraints}.
\begin{equation}
\label{eq:fdm_posterior_jacobian}
P(\fdm \mid N, \NL=0, \Ml) = P(u \mid  N, \NL=0) \, \dfrac{du\left(\fdm, \Ml\right)}{d\fdm}
\end{equation}

\begin{figure}[t]
    \centering
    \includegraphics[width=\columnwidth]{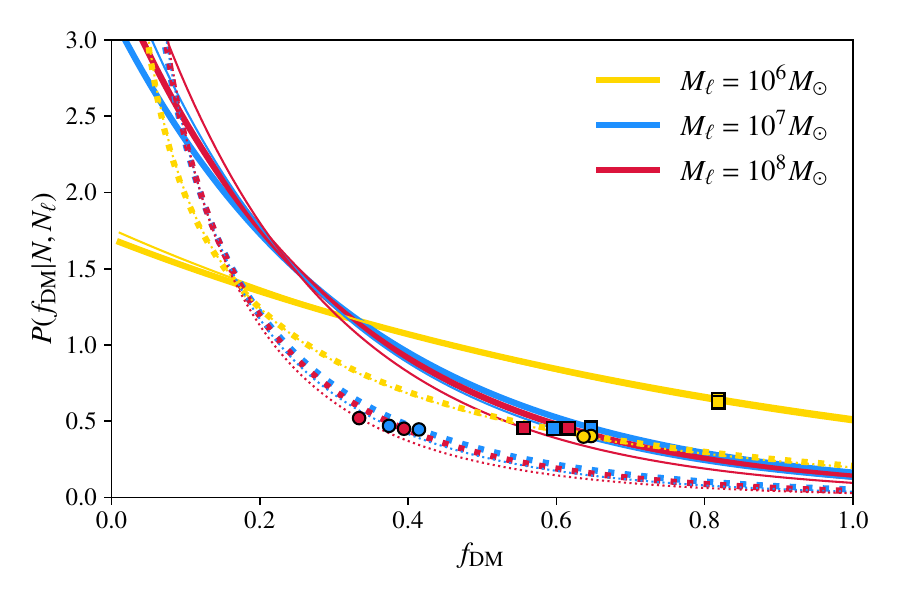}
    \caption{Posterior distribution of the fraction of dark matter $\fdm$ in the form of COs of mass $\Ml$. Solid (dotted) lines and filled squares (circles) respectively show the posteriors and 90\% upper bounds on $\fdm$ assuming flat (Jeffreys) prior in Eq.~(\ref{eq:u_posterior}). We have assumed that the BBH mergers have component masses are distributed according to the ``power law + peak'' model from \cite{abbott2023population}, and are distributed in redshift following \cite{madau2014cosmic} (thick lines) or \cite{belczynski2016first} (thin lines).}
    \label{fig:fdm_posteriors}
\end{figure}

\begin{figure}[t]
    \centering
    \includegraphics[width=\columnwidth]{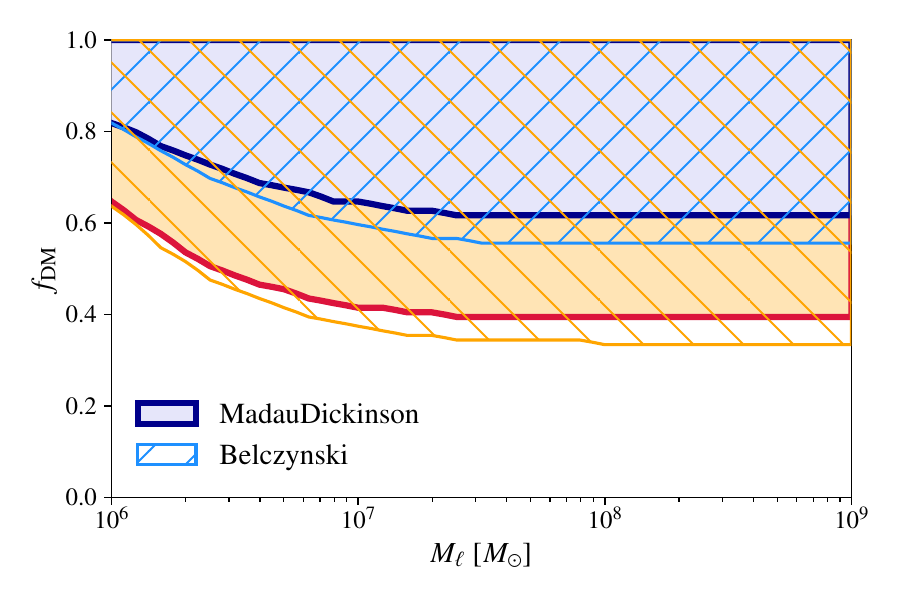}
    \caption{90\% upper bounds on the fraction of dark matter $\fdm$ in the form of COs of mass $\Ml$. Results are shown for two different source redshift distributions (shaded regions vs hatches) and two different priors in Eq.~(\ref{eq:u_posterior}), flat (blue shades), and Jeffreys (red shades).}
    \label{fig:fdm_constraints}
\end{figure}

Figures \ref{fig:fdm_posteriors} and \ref{fig:fdm_constraints}, respectively, show the posteriors and 90\% upper bounds on $\fdm$ for different choices of prior and source redshift distributions. The posteriors show higher support for lower values of $\fdm$ and are better constrained for higher lens masses, being most constraining for $\Ml \sim 10^{9}~\Msun$. The latter is a direct consequence of the lensing fraction being higher for higher $\Ml$. We can constrain $\fdm \lesssim 0.4-0.6$ with current LIGO-Virgo data. We remark that most of the uncertainty arises due to our choice of prior, with little dependence on the assumed redshift distribution.

\begin{figure}[t]
    \centering
    \includegraphics[width=\columnwidth]{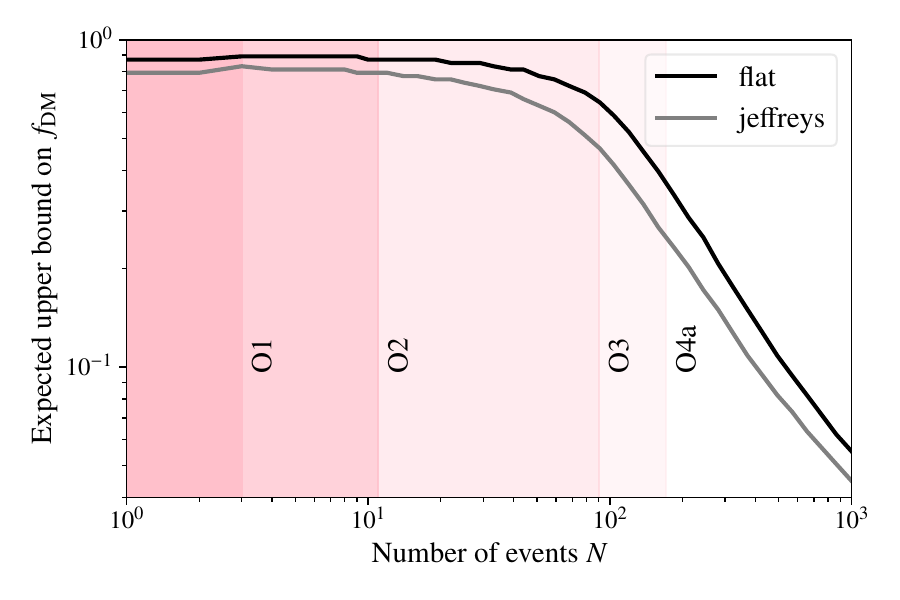}
    \caption{Current and expected 90\% upper bounds on the fraction of dark matter $\fdm$ in the form of COs of mass $\Ml=10^7~\Msun$, as a function of the number of BBH observed mergers with no detected lensed events. The black and grey lines correspond to flat and Jeffreys prior, respectively. We have used a GW source redshift distribution from \cite{madau2014cosmic} for this simulation.}
    \label{fig:fdm_constraints_projected_M1e7}
\end{figure}

If LIGO-Virgo detectors continue to detect GW signals without detecting the presence of strong lensing, or if such detections are confidently associated with non-CO lenses, our constraints will become much tighter. Figure \ref{fig:fdm_constraints_projected_M1e7} shows the projected 90\% upper bounds on $\fdm$ for future observing runs. After $N \sim \mathcal{O}(1/u)$ GW signals have been detected, the upper bounds rapidly drop with further detections, raising our hopes of competing with alternative probes of compact dark matter in the near future.

We remind the reader that these results are based on a lower bound obtained on the lensing fraction using the strongest lens approximation (see Appendix~\ref{sec:multiple_lensing_approx}) and an empty cone optical depth (see Appendix~\ref{sec:strong_lensing_optical_depth}). Consequently, the $\fdm$ constraints are a conservative upper bound, and may, in fact, be slightly tighter.

\section{Conclusion}
\label{sec:conclusion}
Various searches have found no evidence of strong lensing in the $\sim 90$ GW events observed by LIGO and Virgo so far. While this non-detection is curbed by uncertainties, a powerful argument based on time delays between observed events can be used to conclusively rule out the presence of lensing by COs of mass $10^{6}-10^{9}~\Msun$. We have used this non-observation to constrain the fraction of dark matter in the form of such COs.

We used a Bayesian formalism coupled with simulations of strong lensing of GW sources in an inhomogeneous cosmology. If dark matter had been in the form of COs of mass $10^{9}~\Msun$, $\sim$3.5\% of all detected GW events would have been strongly lensed. Non-observation of such lensed signals allows us to rule out $40-60\%$ of dark matter in the form of COs of mass $10^{6}-10^{9}~\Msun$. While these are quite modest when compared to those in the literature \citep{carr2021constraints}, they will only improve with the availability of more data, provided we can conclusively rule out strong lensing from observed data.

The latter expectation may seem too optimistic; current estimates of lensing rates \citep{ng2018precise, oguri2018effect} predict that a detection of strongly lensed event is imminent in the upcoming runs of LIGO-Virgo. In case such a detection is made, do our future predictions on $\fdm$ constraints become void entirely? We think that there is still hope of ruling out compact dark matter lenses in such a case and outline a procedure for doing so.

Firstly, a point lens is known to produce exactly two images separated by a morse phase difference of exactly $\pi/2$. Therefore, if more than two lensed counterparts of the same event are detected, or if only two lensed images are detected, but having a morse phase difference of 0 or $\pi$, they could not have been lensed by a point lens. Our simulations show that galaxy lenses have roughly equal chances of producing exactly 2 or more than 2 images. However, our precise ability to rule out CO lensing from multiple images produced by galaxy/cluster lenses needs to be characterized. This is something that we are investigating now.

Our constraints on $\fdm$ are conservative, and they could be tighter. Part of this can be attributed to the fact that we have used an empty cone optical depth for all $\fdm$, even though the optical depth may be slightly larger in partially filled cones. In addition, our result is conservative because we have neglected multiple lensing. However, through simulations of double lensing across different source redshifts, we find that the error would be smaller than 2.5\%, and ignoring multiple lensing would make our constraints only slightly weaker.

\section*{Acknowledgments}
We thank Anuj Mishra for his careful review of the manuscript and Rajaram Nityananda for guidance regarding inhomogeneous cosmologies. We are also grateful to Srashti Goyal, Loganayagam R., Soummyadip Basak, as well as members of the ICTS Astrophysical Relativity group and LIGO-Virgo-KAGRA collaboration's lensing group for helpful discussions and suggestions. Our research is supported by the Department of Atomic Energy, Government of India, under Project No. RTI4001. The numerical calculations reported in the paper were performed on the Alice computing cluster at ICTS-TIFR.

\appendix

\section{Multiple gravitational lensing}
\label{sec:multiple_lensing_approx}

% --------------------------------------------------------------------------------------------------------------------------------%
\begin{figure}[t]
    \centering
    \includegraphics[width=\columnwidth]{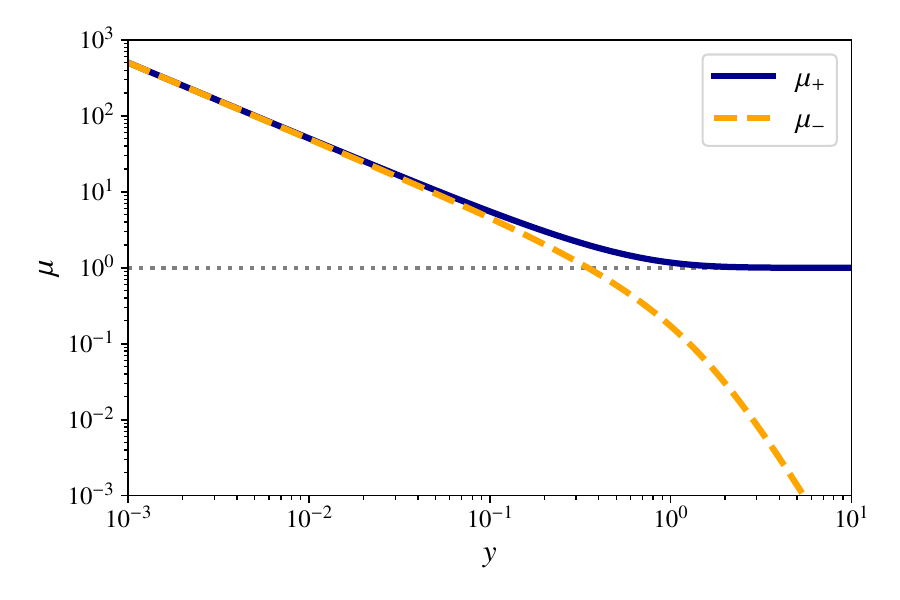}
    \caption{Magnification of images due to a point mass lens as a function of the impact parameters $y$, see Eq.~\eqref{eq:point_lens_mag}.}
    \label{fig:point_lens_mag}
\end{figure}
% --------------------------------------------------------------------------------------------------------------------------------%

We treat multiple lensing as a perturbative phenomenon where the second strongest lens perturbs the solution corresponding to the strongest lens; the third strongest lens perturbs the solution corresponding to the strongest two lenses; and so on. The formula for magnification by a point lens (e.g.,~\cite{meneghetti2021introduction}) gives a natural criterion for deciding the lenses' strength
\begin{equation}
\mu_\pm(y) = \dfrac{1}{2} \pm \dfrac{y^2+2}{2y\sqrt{y^2+4}}.
\label{eq:point_lens_mag}
\end{equation}
As shown in Fig.~\ref{fig:point_lens_mag}, $\mu_\pm$ decrease monotonically with $y$; hence a lens with a smaller $y$ is relatively ``stronger'' than one with a larger $y$.

One may interpret the optical depth $\tau(z_s)/\pi$ as a surface density of lenses projected onto the source plane, where separations are measured in units of the dimensionless impact parameter $y$. If lenses are uniformly distributed on the source plane, the probability distribution of the smallest impact parameter $y_1$ can be derived using elementary considerations of the Poisson distribution.

\begin{figure}[t]
    \centering
    \includegraphics[width=\columnwidth]{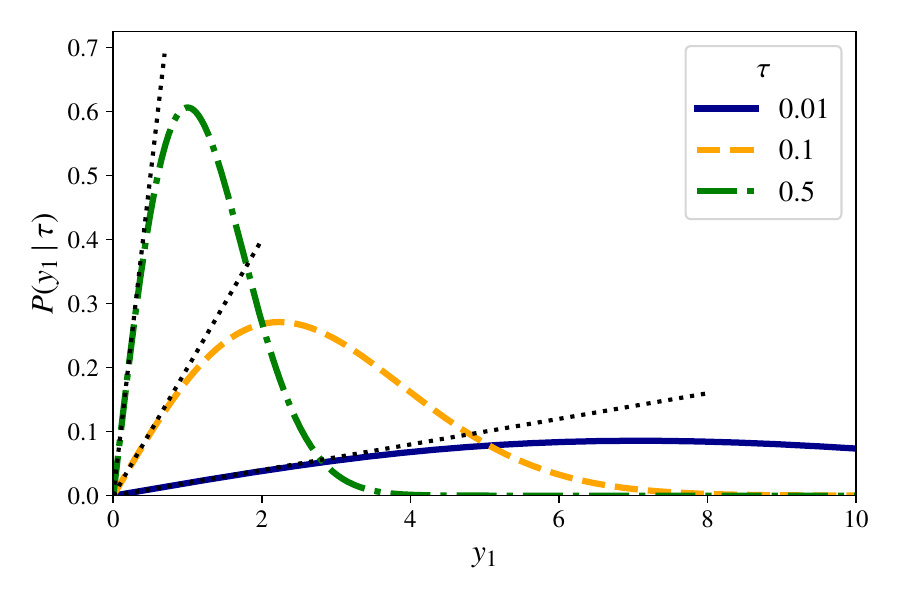}
    \caption{Probability distribution of the strongest lens's impact parameter, Eq.\eqref{eq:P_y1}. The dotted lines show the linear approximation $P(y_1) \propto y_1$. Note that, for large $\tau$, the distribution deviates from the linear approximation even at small values of $y_1$.}
    \label{fig:P_y1}
\end{figure}

The probability $P(y_1)\delta y_1$ of finding the \textit{closest} lens between $[y_1, y_1+\delta y_1] $ is given by the product of probability of finding no lenses in a disk of radius $y_1$, and the probability of finding exactly one lens in an annulus of inner radius $y_1$ and thickness $\delta y_1$. These probabilities are given by
$$P(n \mid \lambda) = \dfrac{\lambda^n e^{-\lambda}}{n!}$$
with $\{n,~\lambda\} = \{0,~\tau(z_s) \, y_1^2 \}$ and $\{1,~2\tau(z_s) \, y_1 \delta y_1\}$ respectively. After taking the small $\delta y_1$ limit, we get
\begin{eqnarray}
P(y_1 \mid z_s) = 2\tau(z_s) \, y_1 \, e^{-\tau(z_s)y_1^2}.
\label{eq:P_y1}
\end{eqnarray}
\cite{fleury2020simple} obtained the same expression using slightly different arguments. We plot this distribution for three values of $\tau$ in Fig.~\ref{fig:P_y1}. While the impact parameters are linearly distributed for small $y_1$, the distribution is highly non-linear at large values of $y_1$, especially for large $\tau$. It is clear that for large $\tau$, one should not assume $P(y_1) \propto y$, as is commonly done in the recent literature on GW lensing.

One can obtain similar distributions for the second smallest, third smallest, etc., impact parameters. For example, the probability distribution function of the second smallest impact parameter $y_2$, conditional on $y_1$, is given by the product of finding no lenses in an annulus of inner and outer radii $y_1,y_2$ respectively, and exactly one lens in a differential annulus of radius $y_2$
\begin{eqnarray}
P(y_2 \mid y_1, z_s) = 2\tau(z_s) \, y_2 \, e^{-\tau(z_s) \, (y_2^2-y_1^2)}.
\label{eq:P_y2}
\end{eqnarray}

If more than one lens is present, no exact analytical formula exists for the magnifications of multiple images. However, several approximation schemes exist in the literature \citep{fleury2020simple}. The strongest lens approximation, first derived by \cite{peacock1986flux}, simply ignores the contribution to magnification from lenses other than the strongest. Therefore, the magnifications are given by $\mu(y_1)$ where $y_1$ follows the distribution given in Eq.\eqref{eq:P_y1}.

Because the optical depth is rather large (Fig.~\ref{fig:z_distrib_optical_depth}), the validity of strongest lens approximation needs to be checked. The most straightforward way is to simulate many random configurations of two compact dark matter lenses and check whether the distributions of magnification $\mu_{1,2}$ and time delays $\Delta T$ --- the only lensing observables relevant to us --- remain similar. We randomly draw the configuration of the first lens by following the same procedure as that stated in Appendix \ref{sec:simulations}, supplement it with Eq.~\eqref{eq:P_y2} for drawing the second lens, and compute $\mu_{1,2}, \Delta T$ using a general purpose numerical solver \textsc{lenstronomy} \citep{birrer2018lenstronomy, birrer2021lenstronomy}.

\begin{figure*}[t]
    \centering
    \includegraphics[width=2\columnwidth]{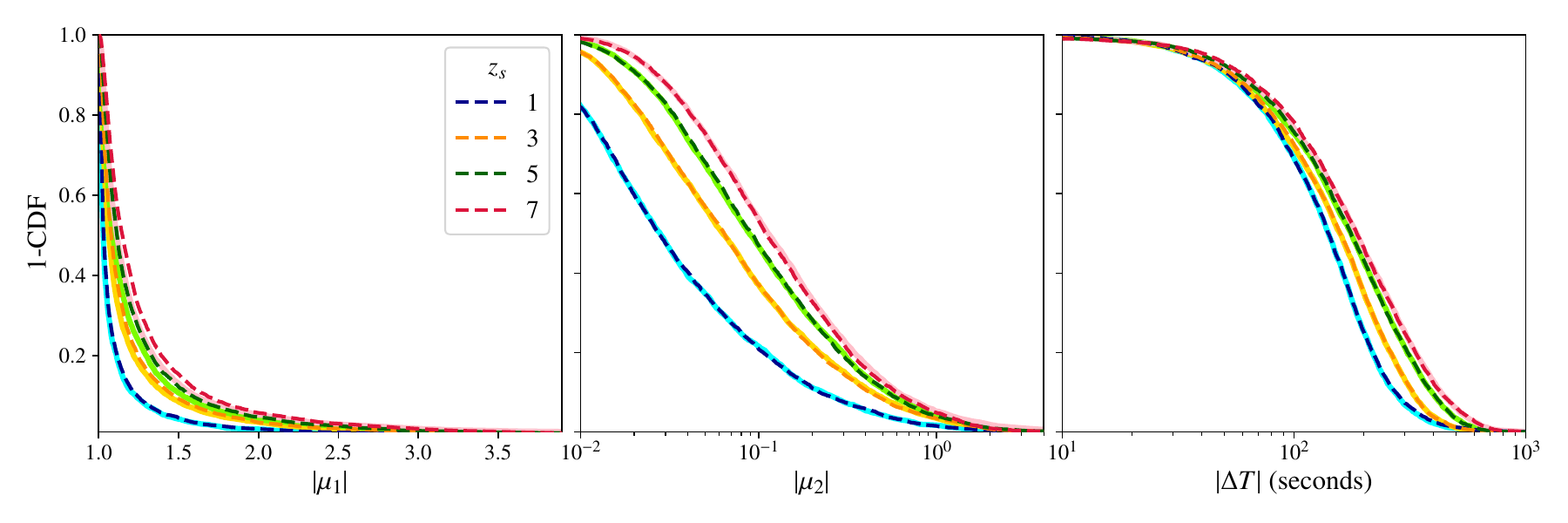}
    \caption{(Reversed) cumulative distributions of magnifications and time delays of the loudest two images from one (lighter solid lines) and two (darker dashed lines) lenses. While the distributions of $\mu_2$ and $\Delta T$ remain more or less unchanged, the distribution of $\mu_1$ is altered slightly due to the presence of a second lens.}
    \label{fig:lenstronomy_double_lensing_distribs}
\end{figure*}

\begin{figure}[t]
    \centering
    \includegraphics[width=\columnwidth]{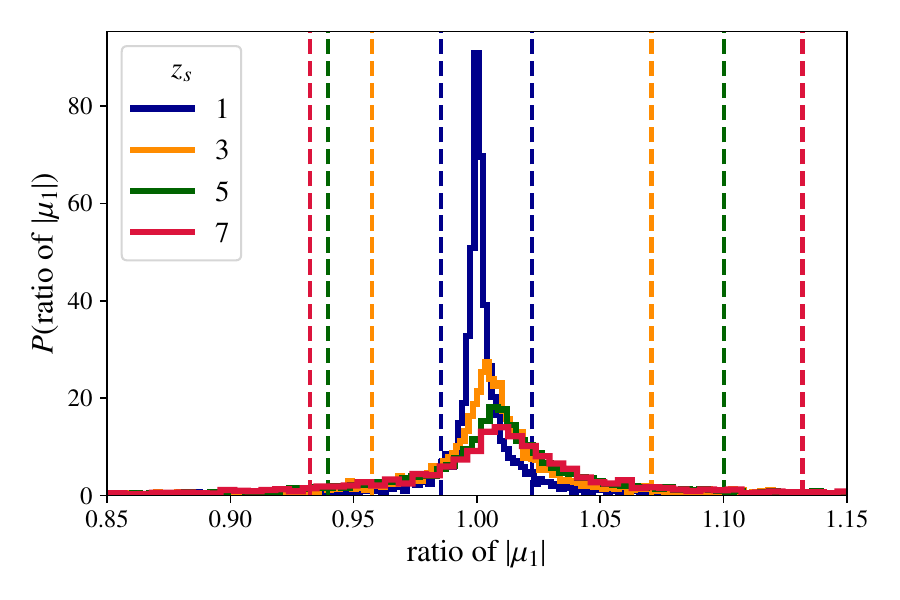}
    \caption{Distribution of ratio of $\mu_1$ calculating assuming double or single lensing. Vertical dashed lines indicate the 10th and 90th percentile bounds.}
    \label{fig:lenstronomy_mu1_ratio_distrib}
\end{figure}

Figure \ref{fig:lenstronomy_double_lensing_distribs} shows the (reversed) cumulative distributions of $\mu_{1,2}$ and $\Delta T$ as obtained from the simulation. It is apparent that though The distributions of $\mu_2$ and $\Delta T$ remain more or less unchanged, double lensing shifts the distribution of $\mu_1$ towards higher values. We also plot the histogram of the ratio of $\mu_1$ calculated with and without the second lens (Fig.~\ref{fig:lenstronomy_mu1_ratio_distrib}). We find that the presence of a second lens can cause the magnification to increase (on average) by $\sim 1\%$, though it may fluctuate by $\sim 7-13\%$ on an individual basis.

Our final goal of this exercise is to estimate the detectable fraction of lensed events. Fluctuations due to double lensing could affect it in two ways: they might change in magnification, which may make some marginal events detectable/not detectable. Because these fluctuations are asymmetric around unity, there will be an overall excess magnification due to double lensing. In the worst case when $z_s=7$, the peak of the magnification ratio in Fig.~\ref{fig:lenstronomy_mu1_ratio_distrib} shifts by at most 1\% (i.e. to 1.01). This will have a similar effect as reducing the signal-to-noise (S/N) threshold for detection by $\sqrt{1.01}$. Our simulations show that with this change, the lensing fraction increases at most by 2.5\% (i.e. by a factor of 1.025). Since the optical depth isn't stupendously large, the effect of three or more lenses is expected to be even more subdominant. It is clear, then, that though multiple lensing is possible, the error incurred by ignoring it is negligible, and would only make our estimates slightly conservative.

\section{Astrophysical simulations of strongly lensed mergers}
\label{sec:simulations}
Here we sketch our methodology for estimating the detectable fraction of strongly lensed events. This requires simulating a population of BBH mergers, computing the effects of strong lensing due to point lenses constituting a fraction $\fdm$ of dark matter, and checking their detectability.

\subsection{Simulating astrophysical sources of GWs}
Since we want to calculate the fraction of lensed events that LIGO-Virgo might have observed in the first three observing runs, the most relevant sources are BBH mergers. Moreover, they are also the most likely sources to be lensed, thanks to the larger detection horizon. We generate a population of $10^8$ BBHs following physically motivated distribution functions for various parameters. We assume that their masses are distributed according to the ``power law + peak'' model inferred from the population of BBHs detected by LIGO-Virgo \citep{abbott2023population}.

While \cite{abbott2023population} also provide a redshift distribution for the binaries, its range is limited to $z\lesssim 1.5$. Since strong lensing can magnify events that would otherwise be out of our horizon, we must simulate a source population far deeper than our estimated horizon redshift for unlensed events \citep{wierda2021beyond}. Hence, instead of using the \cite{abbott2023population} redshift distribution, we assume that the redshift distribution of BBH mergers is consistent with the cosmic star formation history given in \cite{madau2014cosmic}. We find that a maximum redshift of 7 is sufficient to ensure that we don't miss out on potential lensed events from beyond the horizon (Fig.~\ref{fig:z_distrib_optical_depth}).

We assume the BBH to be in quasi-circular orbits with uniformly distributed component spins aligned with the direction of angular momentum. We further assume that they are uniformly distributed in binary phase and polarization, and isotropic in inclination and sky angles. Their arrival times are assigned uniformly within the duration of the observing run.

\subsection{Simulating strong lensing by compact dark matter}
As justified in Appendix~\ref{sec:multiple_lensing_approx}, we use the strongest lens approximation outlined in \cite{peacock1986flux} and \cite{fleury2020simple} to simulate the effects of strong lensing. For each BBH source in our simulated population, we draw a lens redshift $\zl$ according to the differential optical depth (Eq.~\ref{eq:diff_opt_depth})
\begin{equation}
\label{eq:prob_lens_redshift}
P(\zl \mid z_s) \propto \dfrac{d\tau}{d\zl}(\zl, z_s),
\end{equation}
with the normalization fixed using numerical integration. We use an inhomogeneous ``empty cone'' cosmology for all our optical depth calculations irrespective of the value of $\fdm$ (see the discussion in Sec.\ref{sec:strong_lensing_optical_depth}). This causes an underestimation of the lensing probability for small $\fdm$, leading to slightly conservative constraints.

The dimensionless impact parameter $y$ corresponding to the strongest lens is chosen according to Eq.~\eqref{eq:P_y1}. With $\zl$ and $y$ known, the magnifications of the two images are given by Eq.~\eqref{eq:point_lens_mag}, and the time delay between them is given by \cite{takahashi2003wave}
\begin{equation}
\label{eq:point_lens_timedelay}
\dTl = 4 \Ml (1 + \zl) \left[\dfrac{y \sqrt{y^2+4}}{2} + \log{\dfrac{\sqrt{y^2+4} + y}{\sqrt{y^2+4} - y}} \right],
\end{equation}
where $\Ml$ is the mass of the point lens.

\subsection{Checking the detectability of unlensed and lensed GWs}
In principle, almost every source in our simulation is lensed (i.e. has a lens in front of it at redshift $0 < \zl < z_s$, with some impact parameter $y$). However, most of these sources will not be intrinsically loud enough, or may not be magnified enough, to be observable by our GW detectors. To utilize the non-observation of strong lensing, we must compute the fraction of \textit{detectable} lensed events in different observing scenarios.

To keep the analysis simple, we use the optimal S/N that a source would produce at our detectors as our detection statistic. We use the \textsc{IMRPhenomD} \citep{purrer2023imrphenomd} waveform model to generate BBH signals and compute the optimal S/N in the frequency range 20--1024 Hz using the \textsc{pycbc} package~\citep{alex_nitz_2024_10473621}. We use the representative/anticipated PSDs from \cite{adligo-psd-o1-h1, adligo-psd-o1-l1, H1L1V1-psd-O2, H1L1V1-psd-O3O4O5}.

We magnify each source's S/N by multiplying it with ${{|\mu|^{1/2}}}$. Then, following similar detection criteria as \cite{goyal2024rapid}, a source is deemed detectable as a ``lensed'' event if the magnified network S/N of the loudest image crosses a threshold of 8, and that for the fainter image crosses 7. If only the former criterion is satisfied, the event is deemed ``unlensed''. If neither criterion is satisfied, the event is deemed undetectable.

Finally, we count the number of sources satisfying the lensing detection criterion to obtain a Monte-Carlo estimate of the detectable lensing fraction
$u \equiv \Lambda_\ell/\Lambda  \simeq {N}^\mathrm{sim}_\ell/{N}^\mathrm{sim}$, where ${N}^\mathrm{sim}$ is the detected number of simulated events (lensed+unlensed) corresponding to each observing run, and ${N}^\mathrm{sim}_\ell$ is the detected number of lensed events from that simulation. The total lensing fraction weighted over different observing runs given by
\begin{equation}
u = \dfrac{1}{N} \sum_R {N}_R u_R.
\end{equation}
where $N_R$ is the number of detections in run $R$, and $N$ is the total number of detections considered. We take $N_R=3,8,79$ respectively for the first three observing runs \citep{abbott2019gwtc, abbott2021gwtc, abbott2023gwtc, abbott2024gwtc}. For future projections, we take $N_\mathrm{O4} = 81\times2 = 162$ for the fourth observing run by extrapolating from O4a \citep{gracedb-o4}.

\bibliography{macho_go}

\end{document}